 \documentclass[referee]{aa} 
\usepackage{graphicx}
\usepackage{lscape}
\usepackage{txfonts}

\newcommand{\sv}{\% $(100~{\rm nm})^{-1}$}

\newcommand{\dl}{D$_{\lambda}$}
\begin{document}
   \title{
The spectrum of (136199) Eris between 350 and 2350 nm: 
Results with X-Shooter\thanks{Observations made during X-Shooter 
Science Verification, program 60.A-9400(A), PIs: Alvarez-Candal and Mason}}


   \author{A. Alvarez-Candal\inst{1}
          \and
          N. Pinilla-Alonso\inst{2}
          \and
          J. Licandro\inst{3,4}
	  \and
	  J. Cook\inst{2}
          \and
          E. Mason\inst{5}
          \and
          T. Roush\inst{6}
          \and
          D. Cruikshank\inst{6}
          \and
          F. Gourgeot\inst{1}
          \and
          E. Dotto\inst{7}
          \and
          D. Perna\inst{8}
          }

   \institute{European Southern Observatory, Alonso de C\'ordova 3107, Vitacura, Casilla 19001, Santiago 19, Chile\\
              \email{aalvarez@eso.org}
         \and
             NASA Post-Doctoral Program, Resident Research Associated at NASA Ames Research Center, Moffett Field, CA, USA 
         \and
             Instituto de Astrof\'{\i}sica de Canarias, c/V\'{\i}a L\'actea s/n, 38200 La Laguna, Tenerife, Spain
         \and
             Departamento de Astrof\'{\i}sica, Universidad de La Laguna, E-38205 La Laguna, Tenerife, Spain
         \and
             Space Telescope Science Institute, 3700 San Martin Dr., Baltimore, MD 21218
         \and
             NASA Ames Research Center, Moffett Field, CA, USA
         \and
             Istituto Nazionale di Astrofisica (INAF), Osservatorio Astronomico di Roma, Italy
         \and
             Istituto Nazionale di Astrofisica (INAF), Osservatorio Astronomico di Capodimonte, Italy
             }

   \date{Received / Accepted }

  \abstract
   {X-Shooter is the first second-generation instrument for the ESO-Very Large Telescope. It as 
a spectrograph covering the $300-2480$ nm spectral range at once with a high resolving power.
These properties enticed us to observe the well known trans-Neptunian object (136199) Eris 
during the science verification of the instrument.
The target has numerous absorption features in the optical and 
near-infrared domain which has been observed by different authors, showing differences
in their positions and strengths.
}
   {Besides testing the capabilities of X-Shooter to observe minor bodies, 
we attempt at constraining the existence of super-volatiles, e.g., CH$_4$, CO and N$_2$, and in 
particular try to understand the physical-chemical state of the ices on Eris' surface.
   }
   {We observed Eris in the $300-2480$ nm range and compared the newly obtained 
spectra with those available in the literature. We identified several absorption features,
measuring their positions and depth and compare them with those of reflectance of pure methane 
ice obtained from the optical constants of this ice at 30 K to study shifts in their positions and 
find a possible explanation for their origin.
   }
   {We identify several absorption bands in the spectrum all consistent with the presence of CH$_4$ ice. 
We do not identify bands related with
N$_2$ or CO. We measured the central wavelengths of the bands and find variable shifts, with respect
to the spectrum of pure CH$_4$ at 30 K. 
   }
   {Based on these wavelength shifts we confirm the presence of a dilution of CH$_4$ in other ice on the surface of 
Eris and the presence of pure CH$_4$ spatially segregated. 
The comparison of the centers and shapes of these bands 
with previous works suggest that the surface is heterogeneous. 
The absence of the 2160 nm band of N$_2$ can be explained if the surface
temperature is below 35.6 K, the transition temperature between the
alpha and beta phases of this ice.

Our results, including the reanalysis of data published elsewhere, point to an 
heterogeneous surface on Eris.
}
\titlerunning{Eris through the eyes of X-Shooter}


   \maketitle
%

\section{Introduction}

To interpret the surface composition of minor bodies it is necessary to observe
their spectra in the widest possible spectral range.
This is typically achieved by observing the target with 
different instruments, techniques, and calibrations.
This results in a lack of simultaneous datasets and having often to rely on other parameters
(e.g., albedo, photometric magnitudes) to scale the complete spectrum.
Luckily, new instrumentation allows us to overcome some of these problems. One of such
new instruments is  X-Shooter (XS hereafter), which is also the first second-generation instrument 
developed for the ESO-Very Large Telescope (D'Odorico et al. \cite{dodor06}) and is currently 
mounted in its unit 2 (Kueyen).

More detailed discussion of XS is given in Section 2, but, as an overview, XS is a spectrograph
that obtains spectra from 300 to 2480 nm in a single image at high resolving powers (from 3,000
up to 10,000).
This wavelength range sufficiently covers many known ices and silicates.
At the time of its science
verification we decided to test its capabilities by 
observing a well known trans-Neptunian object with absorption features in most of this spectral range:
(136199) Eris (hereafter Eris).

Eris, formerly known as 2003 UB$_{313}$, is one of the largest bodies in the
trans-Neptunian region, with a diameter $<2340$ km (B. Sicardy, personal communication),
and, together with other two thousand-km-sized objects
[(134340) Pluto and (136472) Makemake], shows CH$_4$ absorption bands in the spectrum, 
from the visible to the near infrared, (see Brown et al. \cite{brown05}, 
Licandro et al. \cite{lica06b}). 

Rotational light-curves of Eris have had little success obtaining a reliable rotational period 
(e.g., Carraro et al. \cite{carra06} or Duffard et al. \cite{duffa08}). Roe et al. (\cite{roe08})
proposed a rotational period of 1.08 days. The light-curve seems not related to Eris' shape, 
but probably to an albedo patch. Nevertheless, due to the noise in the data, even if this albedo 
patch indeed exists, it is clear that there are no large albedo heterogeneities.

A close look at the spectroscopic data shows that there exist a subtle heterogeneity. 
Abernathy et al. (\cite{abern09}) compared their spectra with that of Licandro et al. 
(\cite{lica06b}) finding different central wavelengths and different depths for a few 
selected methane absorption bands. Nevertheless, recent results pointed otherwise: 
Tegler et al. (\cite{tegle10}), using their own results, and those of Quirico
and Schmitt (\cite{quir97a}), showed that a single component, CH$_4$ diluted in N$_2$, could
explain the wavelength shifts because those shifts change by increasing with increasing  
wavelength.

The orbit of Eris is
that of a scattered disk object (Gladman et al. \cite{gladm08}).  It has a
perihelion distance of 49.5 AU, and it is currently located at 97 AU and
approaching perihelion. At 97 AU, its equilibrium temperature is $\lesssim30$ K 
as estimated by the Stefan-Boltzman law and assuming conservation
of energy. 
Therefore, based on models of volatiles retention (see Schaller and Brown \cite{schal07}) 
and its similarities with Pluto, Eris' surface is expected to be covered in N$_2$,
CO and CH$_4$ ices. Of these species, only CH$_4$ has been clearly seen on Eris.

Therefore, we requested observing time during XS science verification to make the most of 
its high resolving power to detect features of CO and N$_2$, and to obtain Eris’ spectrum
simultaneously between 300 and 2480 nm.

The article is organized as follows: in Sec. 2 we describe the instrument and the observations. 
In Sec. 3 we present the results
obtained from our analysis, while the discussion and conclusions are presented in Secs. 4 and 5
respectively.

\section{Observation and data reduction}

\subsection{X-Shooter}

The instrument is an {\it echelle} spectrograph that has an almost fixed
spectral setup. The observer can choose between SLIT (and
slit width) or IFU (Integral Field Unit) modes. We used the SLIT mode. The detailed
description of the instrument is available at ESO's 
webpage\footnote{{\tt http://www.eso.org/sci/facilities/paranal/instruments/xshooter/}}.
This spectrograph has the ability to simultaneously obtain data over the entire $300-2480$ nm
spectral range by splitting the incoming
light from the telescope into three beams, each sent to a different arm: ultra-violet---blue (UVB), 
visible (VIS), and near-infrared (NIR). 
Using two dicroichs the light is first sent to the UVB arm, 
then the VIS arm and the remaining light arrives to the NIR arm. The disadvantage 
of this optical path is the high thermal background in the K region of the NIR spectrum.

Each arm operates independently from the others, with the sole exception
of the two CCDs (UVB and VIS)  which are controlled by the same FIERA 
(Fast Imaging Electronic Readout Assembly) controller.
Therefore, if the two exposures finish at the same time, their read-out is made 
sequentially. The read-out mode of the NIR detector 
is totally independent from the others, operated by an IRACE 
(Infra Red Array Control Electronics) controller.

\subsection{The observational setup}

We observed Eris on two nights of the XS Science Verification,
August 14$^{\rm th}$ and September 29$^{\rm th}$, 2009. As mentioned
above we used the SLIT mode selecting the high gain readout mode and 
$2\times1$ binning for the UVB and VIS detectors. The read-out and binning for 
the NIR detector are fixed.
The slit widths were 1.0\arcsec, 0.9\arcsec, and 0.9\arcsec, for the UVB, VIS, and NIR 
arms, respectively, giving a resolving power of about 5,000 per arm. As usual with
NIR observations, we nodded on the slit to remove the sky contribution. 
As the nod is made by the telescope, we finished with two exposures per arm per night.

To remove the solar and telluric signals from Eris spectra, 
we observed the G5 star SA93-101 (Landolt \cite{lando92}) with the same observational setup as
Eris and a similar airmass to minimize effects of differential refraction.
Details of the observations are presented in Table \ref{table1}.
\begin{table*}
\caption{Observational Circumstances. 
We show the exposure time per each arm, t$_{\rm ARM}$.}         
\label{table1} 
\centering                        
\begin{tabular}{c c c c c c c}
\hline\hline                 
Date$^a$ & Airmass$^b$ & Seeing$^b$ (\arcsec)&t$_{\rm UVB}$ (s)&t$_{\rm VIS}$ (s)&t$_{\rm NIR}$ (s)& Star (Airmass)\\ 
\hline                   
2009-08-14UT07:24 & 1.073-1.146 & 0.7-0.8 & $2\times1680$ & $2\times1680$ & $2\times1800$ & SA93-101 (1.113) \\
2009-09-29UT05:50 & 1.065-1.163 & 0.9-1.2 & $2\times1680$ & $2\times1680$ & $2\times1800$ & SA93-101 (1.153) \\
\hline                 
\end{tabular}

\smallskip
$^a$ Beginning of the observation.\\
$^b$ Minimum and maximum values during the span of the observation.
\end{table*}

\subsection{Data reduction}

The data presented here were reduced using the XS pipeline (0.9.4). There are newer versions
of the pipeline, but we have verified, after processing the spectra, 
that no significant improvements are obtained in terms of SNR, therefore we use version 0.9.4.
The pipeline accounts for flat-fielding, wavelength calibration, merging of different orders, 
and further extraction of the spectra.
The whole reduction relies on calibration files taken during daytime
and on a set of static files provided as part of the pipeline release.

The data were wavelength and spatially calibrated by creating a two-dimensional wave map.
This is necessary due to the curvature of the {\it echelle} orders. The map is created 
by taken images of ThAr lamps combined with pinhole masks, for all three arms.
The finale accuracy of the calibration is better than 1 \AA\ through
the complete spectral range.

After careful evaluation we decided not to use the one-dimensional spectra produced by the pipeline,
but instead make our own extraction using a merged two-dimensional spectrum generated during
the process. The extraction was made in a usual way with {\tt apall.iraf}.

\begin{figure}
   \centering
   \includegraphics[width=8.5cm]{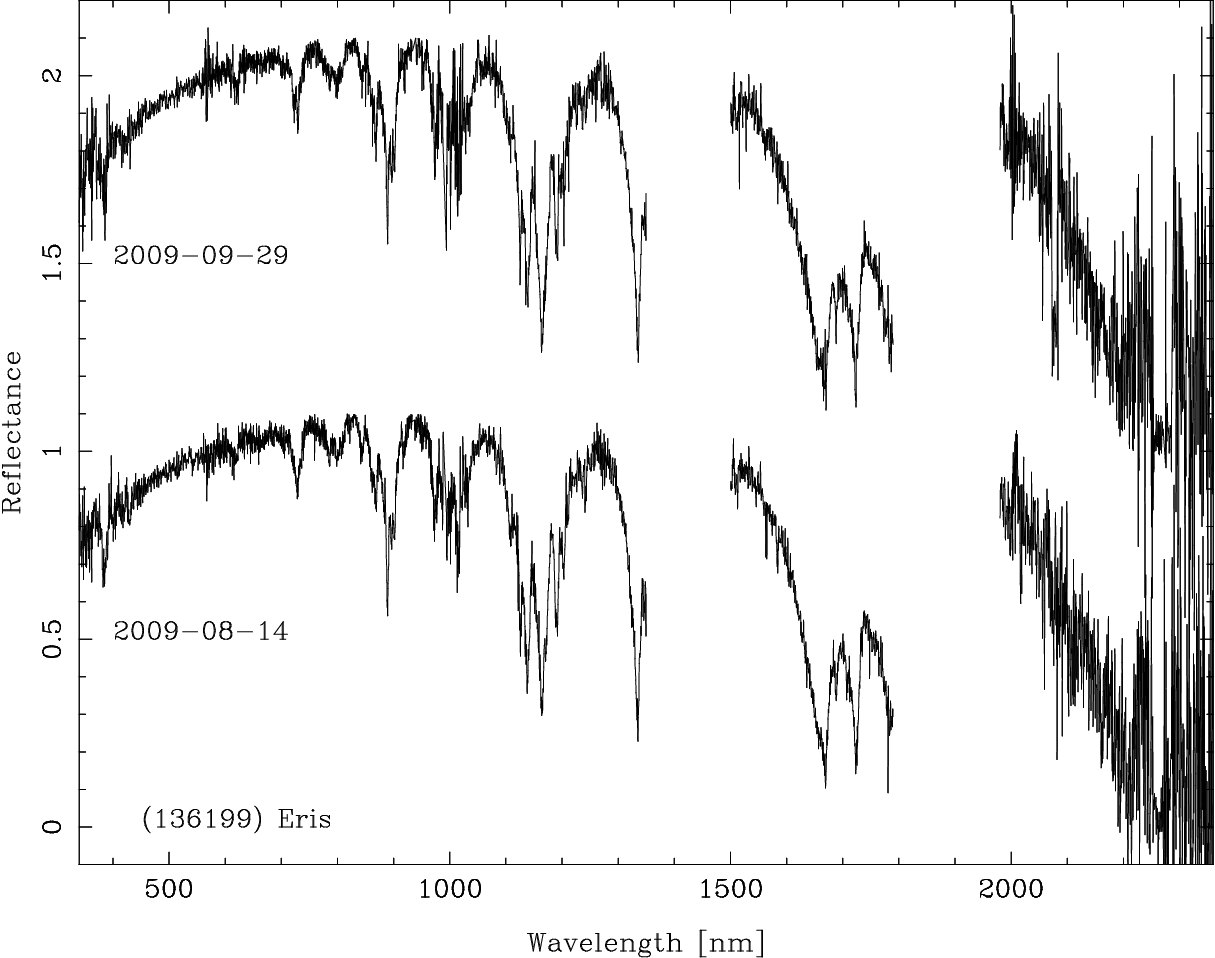}
      \caption{Eris spectra obtained with X-Shooter.
The resolving power is about 5,000 for the spectral
range. The spectra were arbitrarily 
normalized to unity at 600 nm. The spectrum of September 2009 was shifted by 1 in the 
flux scale for clarity. Note that we removed parts with strong atmospheric absorption.}
      \label{fig1}
\end{figure}

Once all spectra were extracted, we divided those of Eris by the corresponding star, used 
as telluric and solar analogue star 
(Table \ref{table1}). 
After the division we removed
the remaining bad pixels from the spectra using a median filtering technique
(as in Alvarez-Candal et al. \cite{alcan07}).
This last step left us
with three separated spectra, one per arm. As there exists a small overlap between their 
spectral ranges, the construction of the final spectrum is straightforward. 
The resulting spectra, between 350 and 2350 nm, normalized to unity at 600 nm, are presented in 
Fig. \ref{fig1}.

A few of the {\it echelle} orders were not well merged, in particular in the visible.
We checked that none of them removed important information, e.g., being the
minimum or the edge of an absorption band, and decided to ignore them.
Two other features at 1576 and 1583 nm, each in a different spectrum,
are artifacts introduced by the star and poor telluric correction, respectively. 
Another feature, at 2160 nm, is due to an order badly merged during the reduction process.
Finally, the feature at about 400 nm is introduced by the star (Alvarez-Candal et al. \cite{alcan08})
and does not represent any compound present on Eris surface. All these features are marked
in Figs. \ref{fig2} to \ref{fig4}.

\section{Spectroscopic Analysis}

\begin{figure} 
	\centering
   \includegraphics[width=8.5cm]{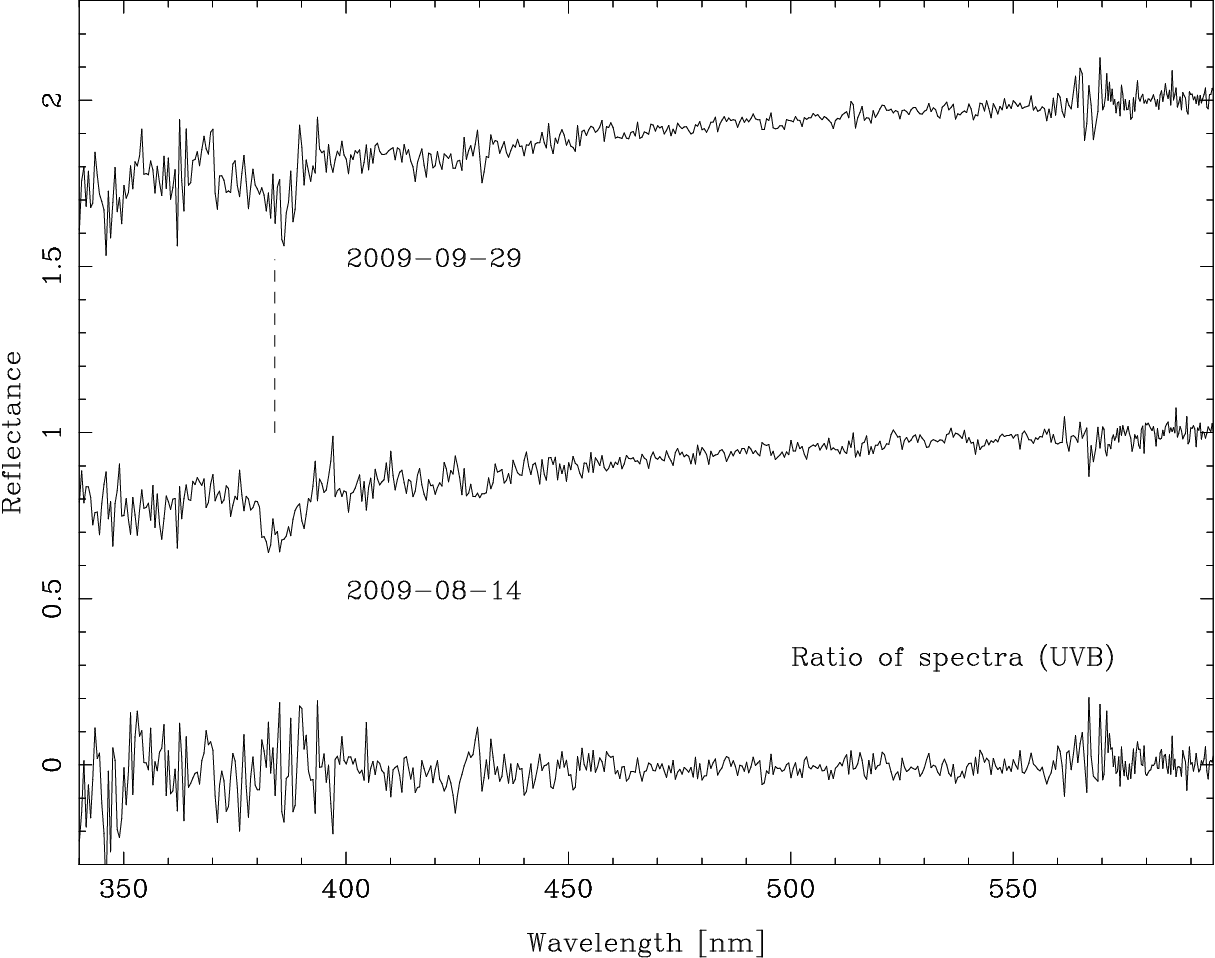}
	\caption{
Ratio of the X-Shooter spectra obtained in the ultra-violet---blue region. 
The artifacts mentioned in the text are marked with a vertical dashed line. 
Normalization and offsets are the same as in Fig. \ref{fig1}. At the bottom it 
is shown the ratio of both spectra, offset by -1 in the scale for clarity. 
}
	\label{fig2} 
\end{figure}

\begin{figure} 
	\centering
   \includegraphics[width=8.5cm]{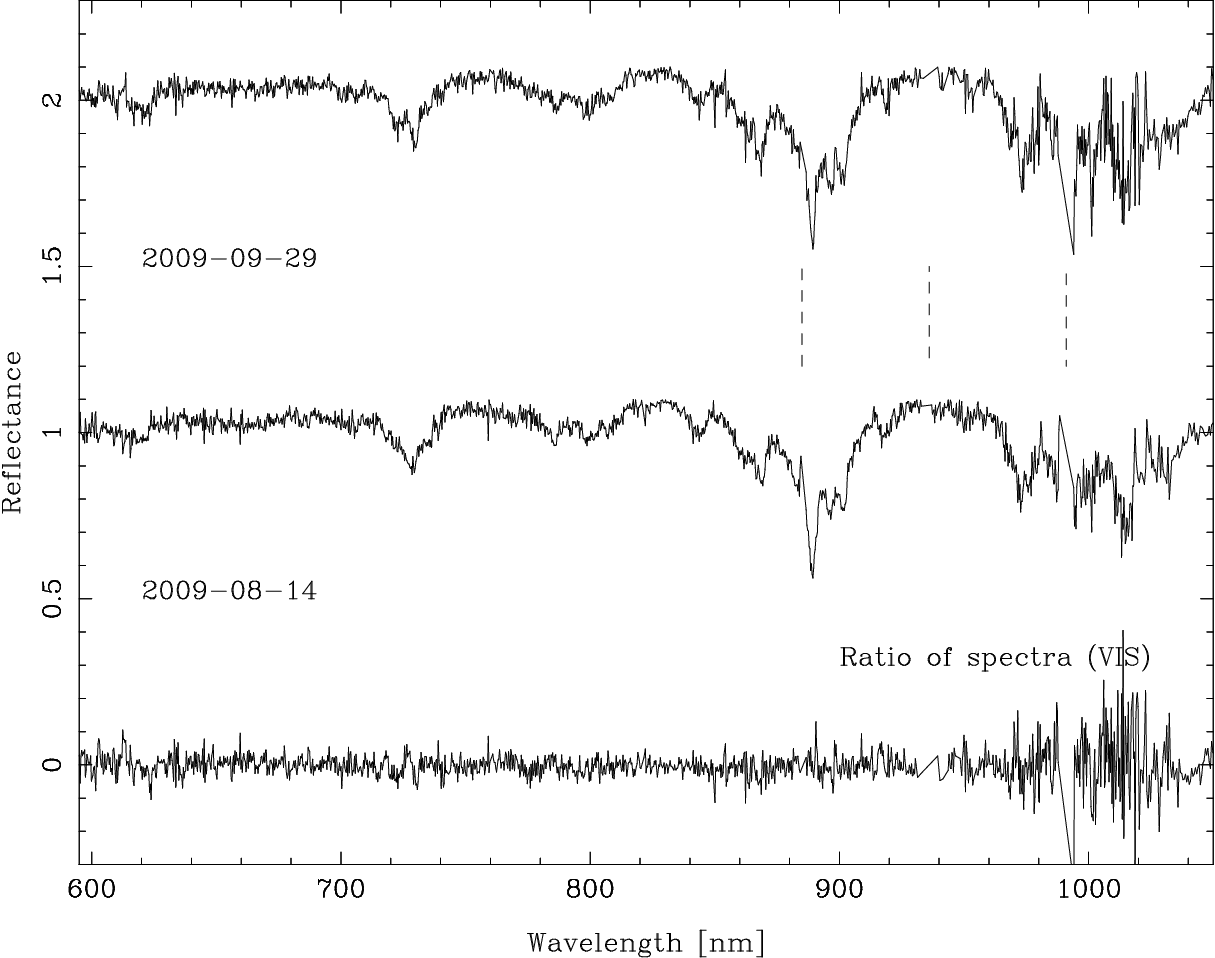}
	\caption{
Same as Fig. \ref{fig2}, but only showing the visible region.
}
	\label{fig3} 
\end{figure}

\begin{figure} 
	\centering
   \includegraphics[width=8.5cm]{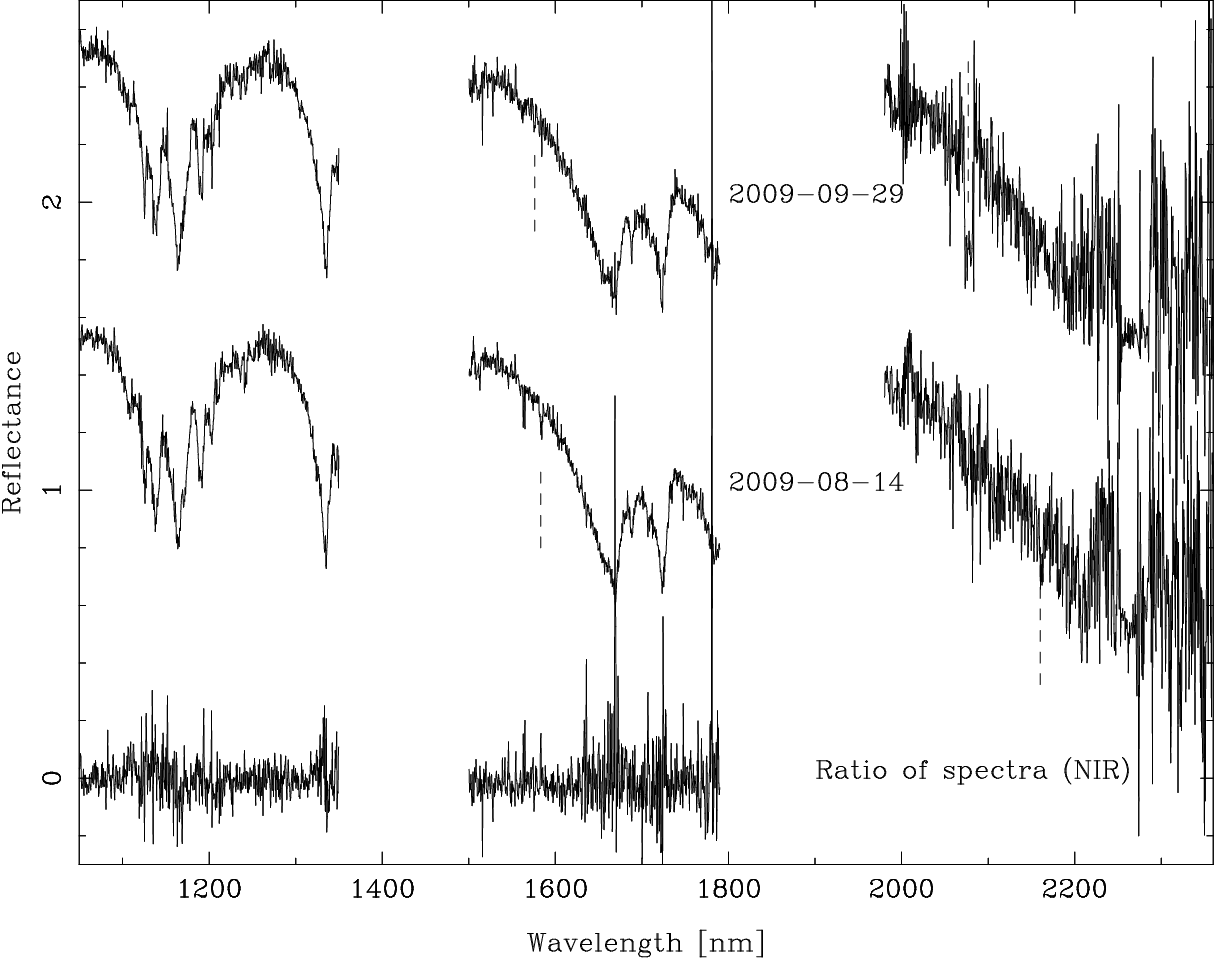}
	\caption{
Same as Fig. \ref{fig2} but showing the near-infrared region. The normalization is the same as in
Fig. \ref{fig1}, but we applied offstes of 0.5 and 1.5 in the flux scale for the August and September
2009 spectra, respectively, for clarity. The ratio of the spectra is not shown in the K region due to
low SNR.
}
	\label{fig4} 
\end{figure}

The spectra presented in Fig. \ref{fig1} are similar throughout the complete wavelength range, 
as shown by the small residues resulting from the calculated ratio of both spectra (Figs. 
\ref{fig2}, \ref{fig3}, and \ref{fig4}).
They both present clear CH$_4$ absorptions from the visible up to the near infrared and
have similar spectral slopes in the visible (see first row of Table \ref{table3} below).

One interesting absorption detected on Eris spectra is located at 620 nm.
This band has been observed in all the giant planets and in Titan's spectrum
and it appears in other previously published spectra of Eris, but has not been discussed.
The band has been assigned to CH$_4$ in its gaseous (Giver \cite{giver78}) and liquid 
(Ramaprasad et al. \cite{ramap78}, Patel et al. \cite{patel80}) phases, in laboratory experiments.
The band strength is $\sim1$ order of magnitude smaller than the 730 nm band, thus, 
its appearance certainly suggests relatively long path-lengths, and likely high concentrations of CH$_4$.

Also, at first glance, these data resemble those already available in the literature, for example:
Licandro et al. (\cite{lica06b}), Alvarez-Candal et al. (\cite{alcan08}) in the visible;
Dumas et al. (\cite{dumas07}), Guilbert et al. (\cite{guilb09}), and Merlin et al. (\cite{merli09})
in the near infrared. These similarities indicate that there are no spectral differences
within the uncertainties of the data, suggesting an {\it a priori} homogeneous surface.
With this in mind, and along with the datasets mentioned above, we will use a combined spectrum
of Eris, obtained as the average of the two XS spectra, unless explicitly mentioned otherwise.

\subsection{Quantitative interpretation of the spectra}
In order to quantify the information in the Eris spectrum we have determined a number of
spectral parameters and these are listed in Table \ref{table3}.

All the measurements presented in the following sections have been made by our team.
We preferred, for the sake of keeping the data as homogeneous as possible, to measure
all quantities rather than use published values.
Along with our Eris data we present values obtained for other Eris spectra, see below, and
 Pluto (column labeled $h$), these last data from Merlin et al. 
(\cite{merli10}, their April 13, 2008 spectrum.).

The Eris spectra we use are: {\it visible} from 
Licandro et al. (\cite{lica06b}), Alvarez-Candal et al. (\cite{alcan08}, their October 2006 spectrum),
and Abernathy et al. (\cite{abern09}, both spectra); and {\it near-infrared} from Dumas et al. (\cite{dumas07}),
Guilbert et al. (\cite{guilb09}), and Merlin et al. (\cite{merli09}, their December 2007 spectrum).

In this paper we use the shorthand term organics to mean the relatively refractory solid material 
consisting of complex macromolecular carbonaceous material that is similar to the insoluble organic matter 
found in most carbonaceous meteorites. This material contains both aromatic and aliphatic hydrocarbons, 
amorphous carbon, and other materials of undetermined structure.
Organics of this general kind characteristically have very low albedo ($\sim0.02-0.06$)
and distinctive red color in the spectral region $300-1000$ nm.

We measured the spectral slope, which provides an indication of the presence of organic materials.
To compute it we fit the continuum with a linear function, ignoring absorption
features, between 600 and 800 nm. The process was repeated many times, each time removing 
randomly points from the datasets. The average value was adopted as the spectral slope and the 
standard deviation as the error. The results are shown in the first row of Table \ref{table3}.
\begin{table*}[]
\caption{Comparison of spectral parameters: Spectral slope in the visible, $S'_{v}$, in \sv,
and percentile depth of the absorption feature, D$_{\lambda}$ at a given wavelength $\lambda$. 
}\label{table3} 
\centering
\begin{tabular}{r c c c c c c c c c}
\hline\hline
Band (nm)&  $a$ (\%)  &   $b$ (\%) &   $c$ (\%) &  $d$ (\%)  &  $e$ (\%)   & $f$ (\%)    & $g$ (\%)   &  $h$ (\%)  \\
\hline
$S'_{v}$ &$ 3.8\pm1.1$&   (\ldots) &$ 3.4\pm0.5$&$ 2.9\pm0.6$&  (\ldots)   &   (\ldots)  &  (\ldots)  &$12.5\pm1.2$\\
\hline
620      &$ 5.3\pm1.0$&   (\ldots) &$ 3.9\pm3.7$&$ 4.7\pm1.0$&  (\ldots)   &   (\ldots)  &  (\ldots)  &  Undet.    \\
730      &$15.8\pm1.8$&$15.0\pm4.6$&$15.7\pm5.2$&$15.1\pm2.0$&  (\ldots)   &   (\ldots)  &  (\ldots)  &$ 5.6\pm1.1$\\
790      &$ 9.7\pm0.7$&$ 8.1\pm3.6$&$ 8.8\pm3.3$&$ 7.2\pm0.8$&  (\ldots)   &   (\ldots)  &  (\ldots)  &  Undet.    \\
840      &$ 5.9\pm1.4$&  Undet.    &$ 8.2\pm5.4$&$ 5.4\pm1.1$&  (\ldots)   &   (\ldots)  &  (\ldots)  & $2.7\pm0.9$\\
870      &$13.2\pm1.8$&$15.1\pm1.4$&$19.0\pm7.5$&$14.1\pm1.1$&  (\ldots)   &   (\ldots)  &  (\ldots)  &$ 3.8\pm3.3$\\
890      &$46.3\pm1.4$&$35.1\pm7.6$&$55.9\pm8.8$&$41.8\pm3.9$&  (\ldots)   &   (\ldots)  &  (\ldots)  &$16.2\pm1.1$\\
1000     &$20.2\pm8.3$& (\ldots)   &  (\ldots)  &  (\ldots)  &  (\ldots)   &   (\ldots)  &  (\ldots)  &  (\ldots)  \\
1015     &$31.2\pm4.3$& (\ldots)   &  (\ldots)  &  (\ldots)  &  (\ldots)   &   (\ldots)  &  (\ldots)  &  (\ldots)  \\
1135     &$39.3\pm2.1$& (\ldots)   &  (\ldots)  &  (\ldots)  &  (\ldots)   &   (\ldots)  &  (\ldots)  &  (\ldots)  \\
1160     &$54.5\pm5.0$& (\ldots)   &  (\ldots)  &  (\ldots)  &  (\ldots)   &   (\ldots)  &  (\ldots)  &  (\ldots)  \\
1190     &$34.4\pm2.0$& (\ldots)   &  (\ldots)  &  (\ldots)  &  (\ldots)   &   (\ldots)  &  (\ldots)  &  (\ldots)  \\
1240     &$ 9.0\pm0.9$& (\ldots)   &  (\ldots)  &  (\ldots)  &  (\ldots)   &   (\ldots)  &  (\ldots)  &  (\ldots)  \\
1335     &$72.2\pm2.6$& (\ldots)   &  (\ldots)  &  (\ldots)  &  (\ldots)   &   (\ldots)  &  (\ldots)  &  (\ldots)  \\
1485     &  (\ldots)  & (\ldots)   &  (\ldots)  &  (\ldots)  &$18.9\pm 5.8$&$32.1\pm 1.7$&$31.1\pm0.9$&$12.9\pm0.9$\\
1670     &$64.4\pm7.3$& (\ldots)   &  (\ldots)  &  (\ldots)  &$61.2\pm 9.5$&$77.7\pm 7.5$&$82.1\pm8.3$&$59.1\pm2.9$\\
1690     &$17.4\pm4.8$& (\ldots)   &  (\ldots)  &  (\ldots)  &$15.5\pm 6.1$&$16.3\pm 6.8$&$14.6\pm2.2$&$ 5.3\pm0.2$\\
1720     &$66.2\pm7.3$& (\ldots)   &  (\ldots)  &  (\ldots)  &$53.6\pm10.9$&$68.3\pm 5.4$&$59.2\pm3.0$&$50.2\pm1.2$\\
1800     &   Undet.   & (\ldots)   &  (\ldots)  &  (\ldots)  &$66.8\pm 4.8$&$72.4\pm13.7$&$70.3\pm3.4$&$56.0\pm0.6$\\
2200     &   Undet.   & (\ldots)   &  (\ldots)  &  (\ldots)  &$89.5\pm29.2$&$81.1\pm18.1$&$60.4\pm8.4$&$50.7\pm1.1$\\
\hline                                                       
\end{tabular}

\smallskip
Undet. = Undetermined ; References: 
($a$) This work; 
($b$) Abernathy et al. (\cite{abern09});
($c$) Licandro et al. (\cite{lica06b});
($d$) Alvarez-Candal et al. (\cite{alcan08});
($e$) Guilbert et al. (\cite{guilb09});
($f$) Merlin et al. (\cite{merli09});
($g$) Dumas et al. (\cite{dumas07});
($h$) Merlin et al. (\cite{merli10}), Pluto
\end{table*}

The depth of an absorption feature is computed following $$D_{\lambda}[\%]=(1-f_{\lambda})\times100,$$
where $f_{\lambda}$ is the normalized reflectance at the wavelength $\lambda$.
This formula gives the percentile value of the absorption depth of a given band and
information about the quantity of the absorbing material, as well as a possible
indication of the path-length traveled by the light. 
To compute it we first normalized the band by fitting a linear function to the band's borders and dividing
the band by that function. Note that this does not change the position of the minimum, while making it easier
to measure $D$ for multiple, overlapping bands (such as the 1690 nm one). The values
are reported in Table \ref{table3}. All values for Eris are compatible to within the errors.

Information about the physical state of the near-surface ices was evaluated by measuring band
shifts between the Eris data, in several small spectral windows, with respect to
synthetic spectra calculated
using Hapke models describing the light reflected from particulate surfaces. We used the optical
constants of pure CH$_4$ at 30 K (Grundy et al. \cite{grund02}), the most likely temperature at Eris 
surface (J. Cook, personal communication), to model the Eris spectrum and evaluated the goodness 
of the fit using a $\chi^2$ criterion. 
Because the CH$_4$ bands at different wavelengths are caused by differing path-lengths,
we used two different grain size as free parameters in the models. The grain sizes range
from tens of micrometers to the scale of meters (for the weakest bands). 
Using such a large range of particle sizes
was necessary because we are only using pure CH$_4$, a more complete modeling 
(beyond the scope of this article) including 
neutral darkening and reddening material would likely result in smaller particle sizes.

The spectral model uses only the optical constants
of CH$_4$ ice from Grundy et al. (\cite{grund02}) at full resolution. 
The optical constants were shifted from $-2$ to $+1$ nm at 0.1 nm intervals.
At each spectral shift, a best fit model spectrum is derived and $\chi^2$ is evaluated. The $\chi^2$ 
measurements are then fit to a quadratic curve to estimate the best fit shift. 
To evaluate the error, we
calculated a correlation matrix around the minimum and added quadratically the uncertainty 
of the wavelength calibration, as mentioned on the corresponding works: 
1 \AA\ for this work's data and Licandro et al. (\cite{lica06b}),
0.3 \AA\ for Abernathy et al. (\cite{abern09}),
and 4 \AA\ for all the rest, as estimated by Merlin et al. (\cite{merli09}).

\begin{figure}
   \centering
   \includegraphics[width=8.5cm]{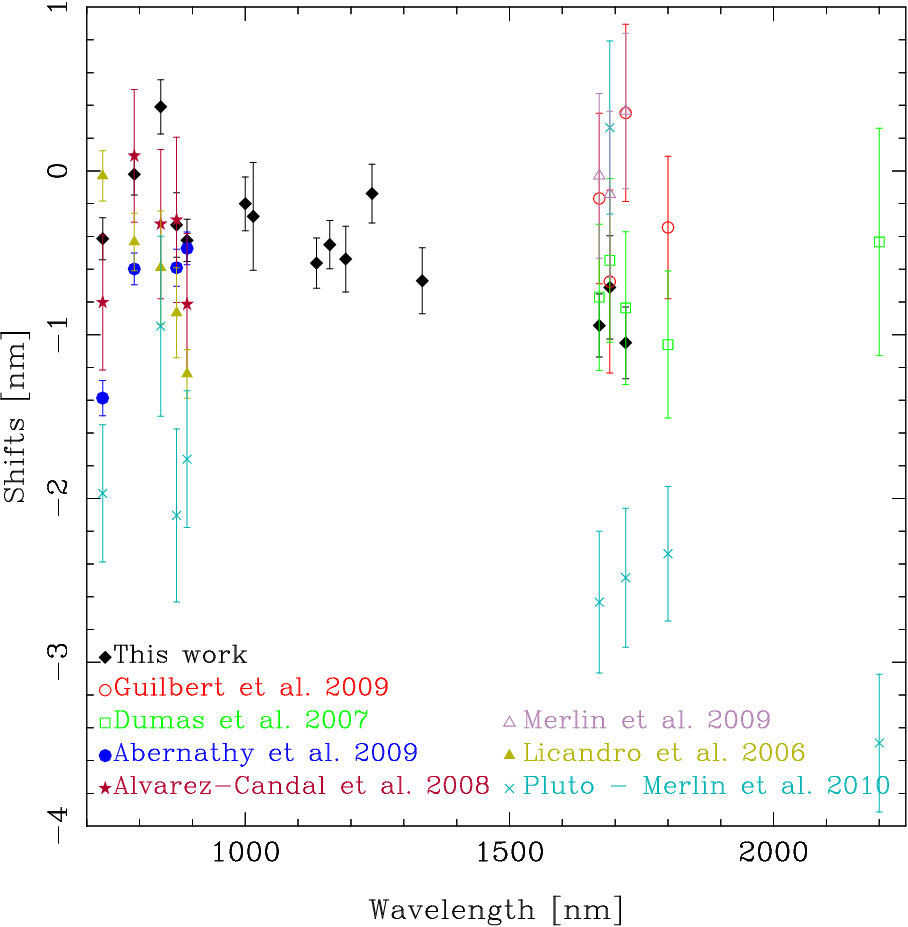}
      \caption{Wavelength shifts for the spectra analyzed in this work.}
      \label{fig6}
\end{figure}

\begin{table*}
\caption{Shifts of CH$_4$ bands, with respect to the spectrum of pure CH$_4$ at 30 K. 
}
\label{table2} 
\centering
\begin{tabular}{r c c c c c c c c }
\hline\hline                 
Band (nm) &$a$ (\AA) & $b$ (\AA)   & $c$ (\AA)   & $d$ (\AA)  & $e$ (\AA)   & $f$ (\AA)  & $g$ (\AA)   & $h$ (\AA)  \\
\hline                          
 730  &$ -4.1\pm1.3$ &$-13.9\pm1.1$&$ -0.3\pm1.6$&$-8.0\pm4.2$&  (\ldots)   &  (\ldots)  &  (\ldots)   &$-19.7\pm4.2$\\
 790  &$ -0.2\pm1.3$ &$ -6.0\pm1.0$&$ -4.3\pm1.7$&$+0.9\pm4.1$&  (\ldots)   &  (\ldots)  &  (\ldots)   &  (\ldots)   \\
 840  &$ +3.9\pm1.6$ &  (\ldots)   &$ -5.9\pm3.4$&$-3.3\pm4.5$&  (\ldots)   &  (\ldots)  &  (\ldots)   &$ -9.5\pm5.4$\\
 870  &$ -3.3\pm2.0$ &$ -5.9\pm1.1$&$ -8.7\pm2.8$&$-3.0\pm5.0$&  (\ldots)   &  (\ldots)  &  (\ldots)   &$-21.0\pm5.2$\\
 890  &$ -4.2\pm1.3$ &$ -4.7\pm1.0$&$-12.4\pm1.5$&$-8.1\pm4.3$&  (\ldots)   &  (\ldots)  &  (\ldots)   &$-17.6\pm4.2$\\
 1000 &$ -2.0\pm1.6$ &  (\ldots)   &  (\ldots)  &  (\ldots)   &  (\ldots)   &  (\ldots)  &  (\ldots)   &   (\ldots)  \\
 1015 &$ -2.8\pm3.3$ &  (\ldots)   &  (\ldots)  &  (\ldots)   &  (\ldots)   &  (\ldots)  &  (\ldots)   &   (\ldots)  \\
 1135 &$ -5.6\pm1.6$ &  (\ldots)   &  (\ldots)  &  (\ldots)   &  (\ldots)   &  (\ldots)  &  (\ldots)   &   (\ldots)  \\
 1160 &$ -4.5\pm1.5$ &  (\ldots)   &  (\ldots)  &  (\ldots)   &  (\ldots)   &  (\ldots)  &  (\ldots)   &   (\ldots)  \\
 1190 &$ -5.4\pm2.0$ &  (\ldots)   &  (\ldots)  &  (\ldots)   &  (\ldots)   &  (\ldots)  &  (\ldots)   &   (\ldots)  \\
 1240 &$ -1.4\pm1.8$ &  (\ldots)   &  (\ldots)  &  (\ldots)   &  (\ldots)   &  (\ldots)  &  (\ldots)   &   (\ldots)  \\
 1335 &$ -6.7\pm2.1$ &  (\ldots)   &  (\ldots)  &  (\ldots)   &  (\ldots)   &  (\ldots)  &  (\ldots)   &   (\ldots)  \\
 1670 &$ -9.4\pm1.9$ &  (\ldots)   &  (\ldots)  &  (\ldots)   &$-1.7\pm5.2$&$-0.3\pm5.0$&$ -7.7\pm4.5$ &$-26.3\pm4.3$\\
 1690 &$ -7.1\pm3.2$ &  (\ldots)   &  (\ldots)  &  (\ldots)   &$-6.8\pm5.6$&$-1.4\pm5.1$&$ -5.5\pm5.0$ &$ +2.6\pm5.3$\\
 1720 &$-10.5\pm2.2$ &  (\ldots)   &  (\ldots)  &  (\ldots)   &$+3.5\pm5.4$&$+3.7\pm4.7$&$ -8.4\pm4.7$ &$-24.8\pm4.2$\\
 1800 &  (\ldots)    &  (\ldots)   &  (\ldots)  &  (\ldots)   &$-3.5\pm4.3$&     Undet. &$-10.6\pm4.5$ &$-23.4\pm4.1$\\
 2200 &  (\ldots)    &  (\ldots)   &  (\ldots)  &  (\ldots)   &    Undet.  &     Undet. &$ -4.3\pm6.9$ &$-34.9\pm4.2$\\
\hline                                                                                                        
\end{tabular}                                                                                                 

\smallskip 
 Undet. = Undetermined ; References:
 ($a$) This work;                                                                                            
 ($b$) Abernathy et al. (\cite{abern09});                                                                    
 ($c$) Licandro et al. (\cite{lica06b});                                                                     
 ($d$) Alvarez-Candal et al. (\cite{alcan08});                                                                
 ($e$) Guilbert et al. (\cite{guilb09});                                                                     
 ($f$) Merlin et al. (\cite{merli09});                                                                       
 ($g$) Dumas et al. (\cite{dumas07});                                                                        
 ($h$) Merlin et al. (\cite{merli10}), Pluto                                                                   
\end{table*}                                                                                                  
Table \ref{table2} lists the measured shifts in wavelength, and Fig. \ref{fig6}
illustrates the relationship between shifts and wavelength.
Eris observations indicate an average           
blue shift of $\sim5$ \AA, while Pluto is $\sim20$ \AA. 
Note that considering all the available datasets, no apparent relation seems to exits. Nevertheless,
if we consider one isolated spectrum, like XS for instance, it is possible to see that the shifts 
increase towards longer wavelengths (black diamonds in Fig. \ref{fig6}), 
as shown by Tegler et al. (\cite{tegle10}) and Licandro et al. (\cite{lica06b}, 
light-brown triangles) but derived from smaller datasets, while
other spectra do not show any tendency, unless subsets of data are used.
Licandro et al.'s band at 890 nm has a blue shifts comparable to that measured for Pluto.
There are four Eris absorption bands which
show redshifts: Alvarez-Candal et al. (\cite{alcan08}) @ 790 nm, this work's spectrum @ 840 nm, 
and the spectra of Guilbert et al. (\cite{guilb09}) and Merlin et al. (\cite{merli09}) @ 1720 
nm, nevertheless all of them are within three sigmas from a null shift.

The 1690 nm absorption feature, that only appears on pure CH$_4$, also shows non zero 
wavelength shifts, this might be due to a temperature difference between the models 
and the actual ice on Eris' surface. Anyhow, the shift is in all cases within three
sigma from the laboratory position.

As mentioned above, we used only pure CH$_4$ at 30 K in the models. Methane ice at this
temperature is at its phase I. During the modeling, we determined that some bands could not be 
fit with pure CH$_4$(I), but could be better explained by a mix of CH$_4$(I) and CH$_4$(II).
This phase of methane ice, CH$_4$(II), occurs at temperatures below 20.4 K. The detailed results and implications are
beyond the scope of the current report.

Using all the CH$_4$ bands in the Eris data and comparison with similar bands in data reported for
Pluto (Merlin et al. \cite{merli10}), we conclude that at least some of the CH$_4$ on Eris appears to 
be diluted in another
material. By analogy to Pluto we assume N$_2$ is the main diluting agent.
\subsection{Qualitative interpretation of the spectra}

Other than the overall blue shift of the bands discussed in the previous section, at first glance
there does not appear to be any systematic behavior of the individual bands analyzed.
Licandro et al. (\cite{lica06b}) suggested that different measured
shifts indicated different levels of dilution of CH$_4$ in a N$_2$ matrix:
Pure CH$_4$ at the bottom, with CH$_4$ concentration decreasing towards the surface. 
The proposed mechanism is based on the fact that N$_2$ condenses at a lower temperature than 
CH$_4$. Therefore, while approaching aphelion CH$_4$ condensed first while N$_2$ was still in
its gaseous phase. As Eris gets farther away from the Sun, its surface temperature decreases
and N$_2$ starts condensing as well, thus the mixing of CH$_4$ in N$_2$ increases.
This was later supported by Merlin et al. (\cite{merli09}), who also added a 
top layer of pure CH$_4$, as a result of the null shift they measured in near-infrared bands.
In contrast, Abernathy et al. (\cite{abern09}), measured blue shifts for five selected absorptions 
and identified a
correlation between the shift and the geometric albedo at the wavelength of maximum band absorption, 
opposite to the behavior reported by Licandro et al. (\cite{lica06b}). They proposed the difference 
was due to a heterogeneous surface.
More recent data presented by Tegler et al. (\cite{tegle10}),
on Pluto and Eris, suggests that this stratigraphic analysis of the data has overlooked the
fact that the blue-shifts increase as CH$_4$ becomes more diluted, i.e., less abundant, in N$_2$.

Our results support Abernathy et al.'s (\cite{abern09}) suggestion: that Eris surface is heterogeneous.
Considering all the data sets, there is no apparent correlation between the shift of central
wavelengths of bands and the supposed depths where they form (deeper bands closer to the
surface, shallower bands deeper on the surface, Fig. \ref{fig7}).
The XS data alone shows increasing blue shifts towards longer wavelength (shallower depths), 
which supports the condensation mechanism proposed in Licandro et al. (\cite{lica06b}), 
but using all the data we cannot confirm this hypothesis. In conclusion,
any stratigraphic analysis should be regarded as local representing the part of the surface
observed during the exposure.


\subsection{Comparison between X-Shooter spectra}

The molecule of methane ice is optically very active. It has four fundamental vibrational modes 
in the infrared between 3000 and 8000 nm, but all of them lead to a huge number of overtones and 
combinations, which can be observed with progressively diminishing absorbances into the 
near-infrared and visible wavelength range (Grundy et al. \cite{grund02}).
We can see several of these bands in our two XS spectra. Although both spectra look very similar and 
display the same bands, some of them display subtle differences as can be seen in Figs. \ref{fig2}, 
\ref{fig3}, and \ref{fig4} (up to 1800 nm where the noise in the spectra does not allow detailed 
comparisons). 

There are several factors that can be responsible for these differences. In pure methane ice, 
differences in the optical path-lengths followed by the scattered light result in different depths 
and widths for the bands. Moreover, as investigated by Grundy et al. (\cite{grund02}), 
temperature changes in the ice produce slight differences in the central peaks of these absorption bands. 
The situation is a bit different when CH$_4$ is diluted in other ices, as it could be for Eris, and 
the shape of the bands is influenced by the concentration of CH$_4$ in the matrix. 

According to the classic treatment of molecules trapped in matrices, the guest CH$_4$ molecules can 
exist as isolated molecules or as clusters of various sizes in the nitrogen matrix (dimer, trimer, etc). 
Quirico and Schmitt (\cite{quir97a}) made a detailed study for the case of methane diluted in nitrogen. 
In the specific case of an isolated CH$_4$ molecule (i.e., which has only N$_2$ molecules as first neighbors), 
the motion of the molecule is a slightly hindered rotation, thus, the CH$_4$ bands show a characteristic 
fine structure that can be seen in comparison with the shape of these bands for pure ice. However, for a 
cluster, the different CH$_4$ molecules interact with each other, the molecular rotational motion is 
perturbed or no longer exist, and the profile of a band can be strongly modified with respect to that of 
the isolated molecule (monomer). Consequently, a CH$_4$ absorption band can be the sum of several different bands 
corresponding to the monomer and other various clusters all present in the sample.
Its dilution alters the shape and central frequency of the bands.

\subsection{Comparison with Pluto}

The immediate comparison for Eris is Pluto: both objects present CH$_4$-dominated spectra, and similar sizes. 
Pluto's surface composition is made of N$_2$ (the major constituent of the surface), CH$_4$, and CO ices.
Eris content of pure CH$_4$ seems higher than in Pluto, 
as witnessed by the deeper band at 1690 nm (and also the rest) and the smaller spectral 
shifts of the absorption bands when compared to pure CH$_4$.
Eris lower visible spectral slope indicates a lower content of organics than on Pluto. 

The other ices, N$_2$ and CO, were observed and identified on Pluto by absorption bands at
2150 and 2350 nm respectively. Thus we would expect to find them on Eris.
Unfortunately, due to the low SNR in the K region of our spectra, we did not find any of them.
Note that an overtone of CO is observed on Pluto at 1579 nm (Dout\'e et al. \cite{doute99}) that
we do not detect. The SNR of our spectra in that region is about 15 while the spectral
resolution is more than five times higher than that reported by Dout\'e et al. (750), therefore
if the CO content, and physical-chemical properties, were similar on Eris, then that absorption should
have been detected.

Even if not directly detected, N$_2$ can be inferred on Eris surface. The shifts measured above are evidence
of the mixture of CH$_4$ with another ice. On Pluto that ice is N$_2$, and probably CO.
The measured blue-shifts for Eris are not as large as those measured in Pluto, 
possibly indicating a lower dilution level of CH4 in N2 than for Pluto.


\section{Discussion}

Comparing the different datasets of Eris we do not find any evidence of major heterogeneities on its surface. 
For instance, the values of spectral slopes and depth of absorption bands presented in 
Table \ref{table3} are all compatible within the uncertainties. 
As a comparison, Pluto's surface has variations in albedo of up to 35 \% (Stern et al. \cite{stern97}). 

The visible slope is indicative of the presence of organics, i.e., the redder the slope, the larger 
the amount of organics and the lower the albedo in the visible. 
In the case of Eris all slopes have a value close to 3 \sv, with error bars in the range of 1 \sv\ 
(see Table \ref{table3}), smaller than that for Pluto (12 \sv), thus pointing to a larger fraction of 
organics on Pluto than on Eris. This is compatible with a higher albedo 
in the visible for Eris (over 0.7, e.g., Stansberry et al. \cite{stans08})
compared to Pluto (averaging 0.6, Young et al. \cite{young01}).

One mechanism to form organic material is the polymerization of carbon bearing molecules, such as CH$_4$.
Laboratory experiments show that long-term irradiation of astrophysical ice mixtures (with the presence 
of simple hydrocarbons and nitrogen) results in the selective loss of hydrogen and the formation of an 
irradiation mantle of carbon residues (Moore et al. \cite{moore83}; Johnson et al \cite{johns84}; 
Strazzulla et al. \cite{straz91}). 
They also show how as the radiation dose increases, more complex polymers, poor in hydrogen content, form 
and an initially neutral-colored and high-albedo ice becomes red as the albedo in the visible decreases. Further 
irradiation gradually reduces the albedo at all wavelengths and, finally, the material becomes very dark, 
neutral in color, and spectrally featureless (Andronico et al. \cite{andro87}; Thompson et al. \cite{thomp87}). 
Since 
TNOs are exposed to ultraviolet solar radiation, the solar wind, and the flux 
of galactic cosmic-rays, it is reasonable to expect that an irradiation mantle with a red color 
could be formed on their surfaces. A 
continuous deposition of fresh volatiles, via sublimation and condensation, acts to keep a neutral
color surface with high albedo.

Assuming the lower visible spectral slope of Eris is due purely to a lower content of organics, then
CH$_4$ on its surface if either younger than that of Pluto, or it is somehow protected against polymerization.
One possible explanation would the recent collapse of Eris' atmosphere, that covered its surface masking the 
organics below.
 
Schaller and Brown (\cite{schal07}) showed, using a simple model of loss of volatiles 
as a function of the surface temperature and size of the objects, that Eris, as well
as Pluto, should have retained CH$_4$, N$_2$, and CO, therefore we expect to detect them. Methane
is easily seen, while N$_2$ can be inferred from the wavelengths shifts measured in the visible.
The case of CO is more complicated, as mentioned above we do not have any detection. But,
building on {\it (i)} Schaller and Brown (\cite{schal07}) results, {\it (ii)} the fact that N$_2$ and CO are
very similar molecules and therefore likely to co-exist (Scott \cite{scott76}), {\it (iii)} CO observation
on Pluto; CO should be present on Eris surface. We will discuss below the possible cause
of the non direct detection of either N$_2$ nor CO.

\begin{figure}
   \centering
   \includegraphics[width=8.5cm]{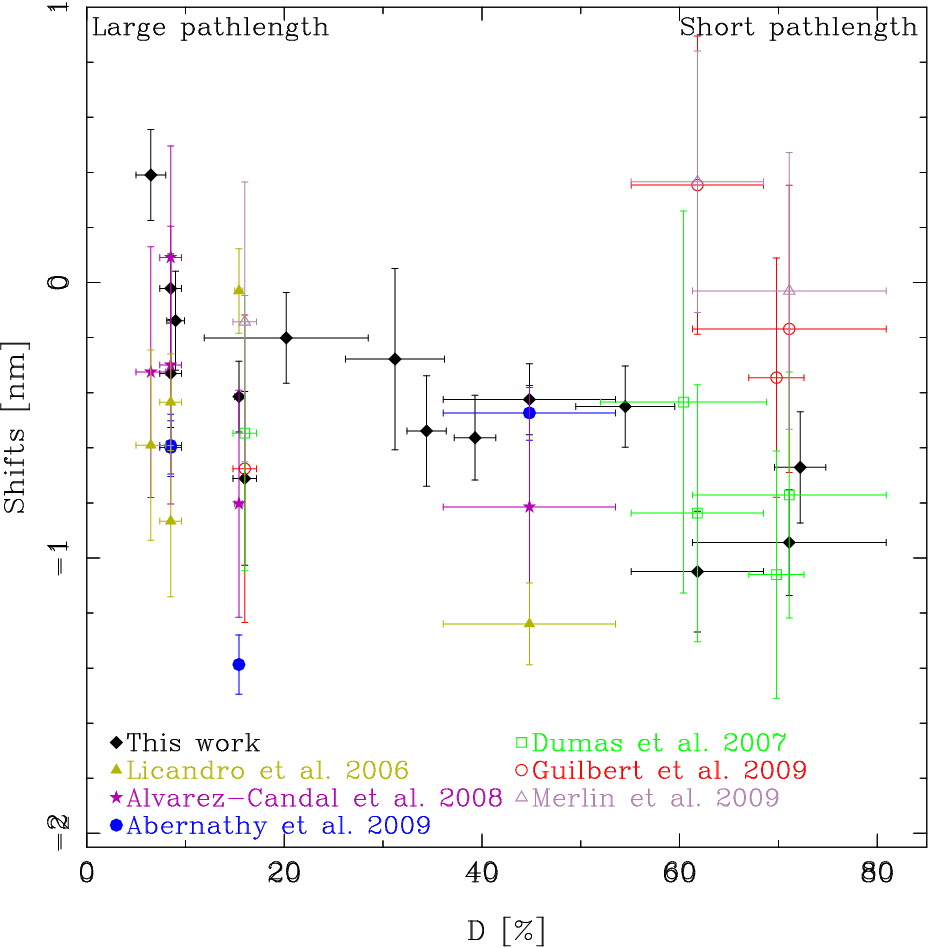}
      \caption{Measured wavelength shifts {\it vs.} averaged \dl. The averaged \dl\ is
the average value as obtained from the Eris entries in Table \ref{table3}, the error bar is
just the standard deviation around the average. \dl\ is used as proxy of real depth in the surface,
as indicated by the legends on the top of the figure. Note that, with exception of XS spectrum,
filled symbols are used for absorption bands in the visible and open ones for bands in the near
infrared ($\lambda>1000$ nm).}
      \label{fig7}
\end{figure}

Inspecting at a more subtle level, wavelength shifts of absorption features,
we find evidence of variable shifts, along one single spectrum and comparing the 
same band in different spectra, which indicate changing mixing ratios of CH$_4$ in another ice, 
probably N$_2$ and/or CO. Different spectra sample different regions on Eris, so, 
assuming that a given absorption is always produced at the same depth, then
most of the layers sampled by the spectra show evidence of heterogeneity (Fig. \ref{fig7}).
Therefore, the surface of Eris is covered by an heterogeneous mix of ices: 
patches of pure CH$_4$ mixed with a dilution of CH$_4$ in N$_2$ and/or CO.
In both cases, the CH$_4$ is so active optically that its features dominate the appearance of the spectrum.

Figure \ref{fig7} was constructed combining the information contained in Tables \ref{table3} and \ref{table2}.
As mentioned above, each absorption band maps a determined depth on Eris' surface, therefore, 
we can picture it as composed of layers. For simplicity
we adopt an average value of \dl, computed from every Eris entry, as a proxy of real depth: topmost 
layers (shorter path-lengths) have larger values of \dl, while deeper layers (large path-lengths) have smaller
values of \dl. 
Given all of the data presented, Fig. 6 shows
variability between datasets suggesting Eris' surface is
heterogeneous. If we consider individual spectra, 
or restricted datasets, relations appear. If we concentrate
only on XS data, black diamonds, the topmost layers show on average the larger blue shifts
and probably higher level of dilution of CH$_4$ in N$_2$ (Brunetto et al. \cite{brune08}), 
while the deepest layer mapped by the 
spectra show purer on average CH$_4$, smaller shifts.

It is important to keep in mind that the relationship between the depth of an absorption band and the path-lengths
traveled by a photon is not a direct one, multiple scattering could account for large path-lengths without the need
to traverse large physical depths.

The uncertainty in Eris rotational period unfortunately precludes any attempt of rotationally resolved
spectroscopy with our current dataset. Considering the Roe et al. (\cite{roe08}) measurement of
$1.08\pm0.02$ days for the rotation of Eris and the 48 days interval between our spectra, 
the indetermination on the rotational
phase will add up to about one day, i.e., almost one rotational period.\\

Why was it not possible to directly detect neither N$_2$ nor CO in Eris spectra?
At the temperature predominating at Eris' orbit (less than 36 K), N$_2$ should be in 
its $\alpha$ state and therefore show an absorption band at 2160 nm, the only band covered by our K region spectra.
However, this band is 
too narrow for the resolution of the spectra and the SNR is insufficient. 
Previous spectra of Eris in the near-infrared (Brown et al. \cite{brown05}
Dumas et al. \cite{dumas07}, Guilbert et al. \cite{guilb09},
Merlin et al. \cite{merli09}) have had better SNR than our XS spectra, in the K region, 
and the resolving power sufficient to detect $\beta$N$_2$ but
have failed to detect it. One possibility is that the majority of the N$_2$, 
if present on Eris' surface, is in the $\alpha$ state.
Quirico and Schmitt (\cite{quir97b}) showed that CO diluted in $\alpha$N$_2$ has a narrow transition band.
Its $0-2$ transition, 2352 nm, is the strongest one and should appear on our data, unfortunately the low SNR
in that region of our spectra does not allow us to impose any constraint on it. However,
as mentioned in Sec. 3.3, its $0-3$ transition at 1579 nm is in a
region with good SNR, but we do not detect it. According to the quoted experimental setup in 
Quirico and Schmitt's work, this transition would need a resolving power of over 12,000 to be observable,
if mixed with $\alpha$N$_2$. This resolution is
twice as large as that of XS, and therefore undetectable with our observational setup.

To detect the features  at 1579 and 2352 nm of CO in a dilution with N$_2$ in $\alpha$-state, 
we would need to have resolving power of more than 10000, which could be achieved with XS 
using its smallest slit width, 0.4\arcsec. 
Given a similar set up and that Eris has $m_v < 17$, we would need several hours
to gather enough signal as to resolve the bands.
The alternative is to wait until Eris enters into a region where the surface temperature rises 
above that of the phase transition from $\alpha$N$_2$ to $\beta$N$_2$, the features get wider and
easier to resolve with less resolution, on the other hand they also decrease in intensity, but a proper 
rebinning of the spectrum should be enough to avoid increasing the exposure times beyond practicality.
However this will not happen until the year 2166.

\section{Conclusions}

Using X-Shooter, a new instrument at the ESO-Very Large Telescope, we obtained
two new spectra of Eris. The spectra were obtained at high resolving power, $\lambda/\Delta\lambda\sim5,000$,
covering the 300 to 2480 nm at once. We compared these datasets with those available
in the literature and with a dataset of Pluto.\\

\noindent
The main results are summarized below:
\begin{itemize}

\item The deeper CH$_4$ absorption bands on Eris indicate a higher content of CH$_4$ than on Pluto. 
This interpretation is also supported by the smaller shift of the CH$_4$ features, from the positions of
pure CH$_4$, when compared to Pluto.

\item Neither N$_2$ or CO are directly detected in our spectra, whereas both have been reported for Pluto.

\item CH$_4$ is probably diluted in another ice, likely N$_2$ and/or CO. This can be inferred from the 
systematic wavelength shift of the CH$_4$ absorption bands in the Eris spectra from all individual observations.

\item We do not see major differences between the Eris spectra, indicating that there are no
large heterogeneities on the regions of its surface sampled by the two observations. 

\item We do observe indications of heterogeneity at a more subtle level. Central wavelengths of
individual CH$_4$ absorption bands have different shifts when independent spectra of Eris are
compared. Considering only the XS data for Eris, there is an indication of an increasing blue
shift with increasing wavelength. This nicely illustrates the advantage the XS design for
simultaneously obtaining data over a broad wavelength range.

\end{itemize}

\vspace{0.5cm}
\begin{acknowledgements}

We would like to thank the X-Shooter team who made it possible that we have these data to work on. 
NPA wants to acknowledge the support from NASA Postdoctoral Program administered by Oak Ridge 
Associated Universities through a contract with NASA.
JL gratefully acknowledges support from the Spanish ``Ministerio de Ciencia e Innovaci\'on'' 
project AYA2008-06202-C03-02. JC acknowledges support of the NPP program.

Also we thank C. Dumas and F. Merlin who kindly made their (reduced) data available to us,
and an anonymous referee for the comments that helped to improve the quality of the
manuscript.

\end{acknowledgements}

\end{document}